\documentclass[fleqn,usenatbib]{mnras}

\usepackage{newtxtext,newtxmath}

\usepackage[T1]{fontenc}

\DeclareRobustCommand{\VAN}[3]{#2}
\let\VANthebibliography\thebibliography
\def\thebibliography{\DeclareRobustCommand{\VAN}[3]{##3}\VANthebibliography}

\usepackage{graphicx}	
\usepackage{amsmath}	
\usepackage[dvipsnames]{xcolor} 
\usepackage{lscape} 
\usepackage{longtable} 
\usepackage{dirtytalk} 

\usepackage{color, colortbl} 
\definecolor{Gray}{gray}{0.9}

\title[X-ray AGN \& SFR in Nearby Galaxies]{The Relationship between the Incidence of X-ray selected AGN in Nearby Galaxies \& Star-formation Rate}

\author[K. L. Birchall et al.]{
Keir L. Birchall$^1$,\thanks{E-mail: klb69@leicester.ac.uk}
M. G. Watson$^1$,
J. Aird$^{1,2}$,
and R. L. C. Starling$^1$
\\
$^1$School of Physics \& Astronomy, University of Leicester, University Road, Leicester LE1 7RH, UK\\
$^2$Institute for Astronomy, University of Edinburgh, Royal Observatory, Edinburgh EH9 3HJ, UK\\
}

\date{Accepted XXX. Received YYY; in original form ZZZ}

\pubyear{2023}

\begin{document}

\label{firstpage}
\pagerange{\pageref{firstpage}--\pageref{lastpage}}
\maketitle


\begin{abstract}
We present the identification and analysis of an X-ray selected AGN sample that lie within the local ($z < 0.35$) galaxy population. From a parent sample of 22,079 MPA-JHU (based on SDSS DR8) galaxies, we identified 917 galaxies with central, excess X-ray emission (from 3XMM-DR7) likely originating from an AGN. 
We measured the host galaxies' star formation rates and classified them as either star-forming or quiescent based on their position relative to main sequence of star formation. 
Only 72\% of the X-ray selected sample were identified as AGN using BPT selection; this technique is much less effective in quiescent hosts, only identifying 50\% of the X-ray AGN.
We also calculated the growth rates of the black holes powering these AGN in terms of their specific accretion rate ($\propto \mathrm{L_X/M_*}$) and found quiescent galaxies, on average, accrete at a lower rate than star-forming galaxies. 
Finally, we measured the sensitivity function of 3XMM so we could correct for observational bias and construct probability distributions as a function of accretion rate. AGN were found in galaxies across the full range of star formation rates ($\log_{10} \mathrm{SFR/M_\odot\ yr^{-1}} = -3\ \mathrm{to}\ 2$) in both star-forming and quiescent galaxies. The incidence of AGN was enhanced by a factor 2 (at a 3.5$\sigma$ significance) in star-forming galaxies compared to quiescent galaxies of equivalent stellar mass and redshift, but we also found a significant population of AGN hosted by quiescent galaxies.

\end{abstract}



\section{Introduction}
Galaxies grow by forming stars, a process driven by the availability of cold gas. This material is also thought to fuel the growth of SMBHs \citep{AlexanderHickox12}. However, a lot of uncertainty remains over exactly how the growth of galaxies and their central SMBHs co-evolve over cosmic time. Large scale surveys of AGN activity across the electromagnetic spectrum have been employed to shed light on the nature of this relationship \citep[e.g.][and many others]{Aird18, Mullaney12a, Rosario13, Smolcic2017, ReinesGreenGeha13}. By exploring how the incidence of AGN changes with the SFR of its host galaxy several key pieces of evidence have been uncovered that suggest the growth of SMBHs and their host galaxies are connected.\\

One of the most well-constrained correlations highlighting the relationship between SMBH and galaxy growth was discovered through including SMBH mass measurements from the nearby Universe. Using a mixture of ground and space-based observations, \cite{Magorrian98} measured the dynamical masses of a sample of 32 SMBHs and uncovered a relationship between a SMBH's mass and the luminosity of the host galaxy's classical bulge. Soon after, similarly strong correlations between related bulge quantities like mass and velocity dispersion \citep[e.g.][]{FerrareseMerritt00, Gebhardt00, Kormendy2013, Greene2020} were found. These results imply the existence of historic, correlated SMBH-galaxy growth.  \\

Further compelling evidence for this relationship has been found by comparing the average properties from large samples. For example, \cite{MadauDickinson14} compiled IR and UV measurements from dozens of previous studies. With this data they measured the cosmic star formation history (SFH) out to $z \sim 6$. They found that, from $z = 6$, the density of star formation in the Universe increased as galactic structures concentrated gas allowing stars to form rapidly. It then peaked at around $z = 1.5$ - $2$, before declining to about a tenth of the peak in the present-day Universe. They then compared this to several measurements of black hole accretion rate density (BHARD), derived from luminosity functions measured in hard X-rays \citep{Aird10} and the infrared \citep{Shankar09, Delvecchio14}. They found a strong agreement between the SFH and BHARD distributions. Thus, it became increasingly clear that there exists a strong connection between star formation and SMBH growth. \\ 

When directly attempting to connect the AGN activity with the host galaxy's SFR, however, the results are not clear. On the one hand, numerous studies found that the average SFR of AGN-hosting galaxies increased out to at least $z \approx 3$ \citep[e.g.][]{Harrison12, Mullaney12b, Rosario12, Rosario13}, and that the SFR is found to tightly correlate with the average AGN luminosity \citep[e.g.][]{Mullaney12a, Chen13}. On the other hand, it was found that AGN at a fixed X-ray luminosity can have a broad range of SFRs \citep[e.g.][]{Alexander05b, Mullaney10} - in some cases covering up to 5 orders of magnitude \citep{Rafferty11}. \\

As the galaxy population evolves through cosmic time, there is a significant change in the star-forming composition of the overall galaxy population. Since $z = 2$, there has been a significant build up of the quiescent galaxy population, particularly in higher mass galaxies \citep{Brammer11, Tomczak14,Barro17}, which highlights a significant amount of star-formation quenching in this period. This decline in the density of star formation across recent cosmic time is thought to be driven by a decreasing density of molecular gas \citep{Popping12,Maeda17}. A concurrent decline in the AGN activity of star-forming galaxies could imply a relationship between star-forming activity and black hole fuelling. \cite{KH09} described this kind of activity through their "feast and famine" fuelling model. By analysing the Eddington ratio distributions of a sample of optically-selected SDSS AGN they were able to demonstrate the existence of two distinct populations of AGN implying there were different regimes of black holes growth. The "feast" mode is associated with galaxies containing significant amounts of star formation in their central regions. The large amounts of cold gas required for star formation fuels black hole growth. The "famine" mode is associated with galaxies hosting older stellar populations. In this case, black hole growth is regulated by the rate at which stars lose their mass. \\

Whilst the optical selection method used in \cite{KH09} produced an incomplete sample \citep{Jones16}, this relationship between star formation and black hole accretion has been observed in more recent studies with more complete samples. For example, \cite{Aird19} explores the effect of star formation on a sample of AGN from the CANDELS survey ($0.1 \leq z \leq 4$), selected using Chandra X-ray data. After applying observational corrections to this sample, they find evidence of an SFR-dependent fuelling mechanism reflective of the model proposed in \cite{KH09}. On the other hand, \cite{Torbaniuk2021} explored this relationship on a sample on AGN at $z < 0.33$ using data from SDSS DR8 and 3XMM DR8. They found twice as many AGN in star-forming galaxies compared to quiescent hosts. They then corrected their sample for observational incompleteness using 3XMM upper limits to study the intrinsic accretion rate distribution in this region. They found systematically larger accretion rates in star-forming galaxies across all stellar masses. Their results imply that both black hole accretion and star formation are fuelled by a common source. \\

In this paper, we explore the star-forming properties of galaxies in the nearby Universe ($z < 0.35$) and investigate how this affects the likelihood of finding an X-ray selected AGN. For this analysis, we use techniques developed in our previous work \citep{Birchall20, Birchall22}. In these papers we investigated the relationship between host galaxy properties and the incidence of X-ray selected AGN in dwarf galaxies, and then in the full galaxy population. Our galaxies were taken from the MPA-JHU (based on SDSS DR8) and X-ray information from 3XMM DR7. We used upper limits to correct our sample for observational incompleteness and used this to measure the probability of finding AGN activity in a galaxy of a given mass and redshift. 

This paper is structured as follows. First, we describe how our AGN sample was constructed (§\ref{sec:data}). Second, we present our approach to classifying the star-forming properties of the galaxies in our sample, making sure to account for the effect of changing stellar mass and redshift (§\ref{sec:SF_Classification}). Then we assess the agreement between our X-ray selected sample and BPT selection, and how star-forming activity affects that classification (§\ref{sec:BPT}). After that, we investigate the rate at which the black holes powering our AGN are accreting material and explore how this changes with its star formation classification (§\ref{sec:sBHAR}). With this data we then measure the probability of finding AGN activity as a function of accretion rate distributions (§\ref{sec:Prob_Dists}) and use them to calculate how the fraction of galaxies that host AGN varies with stellar mass, redshift and star formation rate (§\ref{sec:fractions}). Throughout, we assume Friedmann-Robertson-Walker cosmology: $\Omega = 0.3$, $\Lambda = 0.7$ and $H_0 = 70\ \mathrm{km\ s^{-1}\ Mpc^{-1}}$.

\section{Data \& Sample Selection}
\label{sec:data}
To define a parent sample of local galaxies for this study we used the MPA-JHU catalogue\footnote{Available at \href{http://www.mpa-garching.mpg.de/SDSS/DR7/}{http://www.mpa-garching.mpg.de/SDSS/DR7/}}. This catalogue is based on the Sloan Digital Sky Survey Data Release 8 (SDSS DR8), and includes stellar masses, star formation rates (SFRs) and emission line fluxes for 1,472,583 objects classified as galaxies by the SDSS pipeline. Whilst the MPA-JHU catalogue is formally deprecated by the SDSS, we found the alternatives (Wisconsin, Portsmouth and Granada\footnote{For more information about these catalogues, visit the \href{https://www.sdss.org/dr16/spectro/galaxy/}{\textcolor{blue}{SDSS galaxy properties page}}}) to be insufficient for our purposes. See \cite{Birchall22} for more details on this analysis. \\

The X-ray data comes from the 3XMM DR7 catalogue \citep{Rosen16}. It is based on 9,710 pointed observations with the \emph{XMM-Newton} EPIC cameras in the $\sim 0.2 - 12 \ \mathrm{keV}$ energy range.  We use fluxes in the  $2 - 12\ \mathrm{keV}$ range, by summing the $2 - 4.5\ \mathrm{keV}$ and $4.5 - 12\ \mathrm{keV}$ bands, then converted them to luminosities using the MPA-JHU redshifts, assuming $\Gamma = 1.7$. We used this catalogue, instead of 4XMM, because this was the data used in \cite{Birchall22} and the analysis required the use of comprehensive upper limits data from Flix\footnote{Found at \href{https://www.ledas.ac.uk/flix/flix_dr7.html}{https://www.ledas.ac.uk/flix/flix\_dr7.html}} \citep{Carrera07}. At the time of performing this analysis, 3XMM DR7 was the most recent version of the serendipitous sky survey available with this infrastructure. See \cite{Birchall20} for a more detailed description of both these catalogues.\\

We use the ARCHES cross-correlation tool, \texttt{xmatch} \citep{Pineau17} to match these catalogues. It is an astronomical matching tool which can identify the counterparts of one catalogue to multiple others and compute the probabilities of associations using background sources and positional errors. We broke down the data into individual XMM fields and matched SDSS and X-ray object therein. Using a 90\% probability of association as the matching threshold left us with a well-matched sample of 1,559 X-ray emitting galaxies. See \cite{Birchall22} for more details on the matching process. \\

Our AGN X-ray selection technique was developed in \cite{Birchall20}. First, we modelled the combined emission coming from other X-ray emitting sources - X-ray binary stars ($L_\mathrm{Gas}$, based on the \cite{Mineo12b} model) and hot gas emission ($L_\mathrm{XRB}$, based on the \cite{Lehmer16} model) - and compared it to the observed X-ray luminosity. After we calculated these contributions, we summed them and compared this quantity to the observed X-ray luminosity, $L_{X,Obs}$. Any object that met or exceeded the following criterion was considered to be an AGN, 
\begin{equation}
\label{eq:excessCriterion}
\frac{L_{\mathrm{X,Obs}}}{L_{\mathrm{XRB}} + L_{\mathrm{Gas}}} \geq 3
\end{equation}

Using this criterion, 949 X-ray emitting galaxies were classified as AGN. For the purposes of measuring an accurate AGN fraction we had to ensure we had a statistically complete sample of galaxies above a given stellar mass limit. As in our previous work, we adopt a redshift-dependent stellar mass limit for our sample corresponding the mass above which 90\% of galaxies lie in narrow redshift bins ($\Delta z \sim 0.05$). This results in the removal of 32 objects. See \cite{Birchall22} for full details.

\section{Star-forming Classification}
\label{sec:SF_Classification}
The galaxy population in the nearby Universe is strongly bimodal: there are star-forming galaxies that lie close to or above the main sequence \citep{Noeske07,Elbaz11} and quiescent galaxies that lie below. \cite{Salim07} found that for a large sample of star-forming SDSS galaxies, the main sequence of star formation at a given redshift is described by a power law, $\mathrm{SFR} \propto \mathrm{Mass}^{0.65}$. However, given the relative amount of star formation has changed dramatically through cosmic time \citep[See][among many others]{MadauDickinson14} the exact form of the main sequence relation will evolve with redshift. In this section, we outline the process by which we classify galaxies as star-forming or quiescent using their SFRs relative to this evolving main sequence relation. 

\subsection{Star-forming or Quiescent}
\label{subsec:SF_or_Q}

Our method for splitting the sample into different star-forming classes is based on \cite{Moustakas13}. They calculate a quantity referred to as the "rotated SFR", $\mathrm{SFR_{rot}}$, which attempts to account for the effect of changing stellar mass on SFR. Thus $\mathrm{SFR_{rot}}$ has the form, 
 \begin{equation}
     \log_{10}(\mathrm{SFR_{rot}}) = \log_{10}(\mathrm{SFR}) - 0.65(\log_{10} M_\mathrm{*} - 10)
     \label{eq:rotated_SFR}
 \end{equation} 
 
 where SFR is in units of $\mathrm{M_\odot\ yr^{-1}}$ and stellar mass ($M_\mathrm{*}$) is in units of $\mathrm{M_\odot}$. We plotted histograms of $\mathrm{SFR_{rot}}$ binned by redshift to produce distributions of this mass-independent SFR for the underlying galaxy population. This then allowed us to identify the $\mathrm{SFR_{rot}}$ corresponding to the local minimum between the star-forming and quiescent peaks in each redshift bin. Fitting a straight line to the change of $\mathrm{SFR_{rot}}$ minima with redshift produced an appropriately normalised equation which could be used to split the whole galaxy population into star-forming and quiescent, $\mathrm{SFR_{SF/Q}}$. It took the form,
\begin{equation}
    \log_{10} (\mathrm{SFR_{SF/Q}}) = \log_{10}(\mathrm{SFR}) - 0.65(\log_{10} M_\mathrm{*} - 10) + \frac{z - 0.79}{0.77}
    \label{eq:SF_Split}
\end{equation}

\begin{figure*}
	\centering
     	\includegraphics[width=.85\paperwidth]{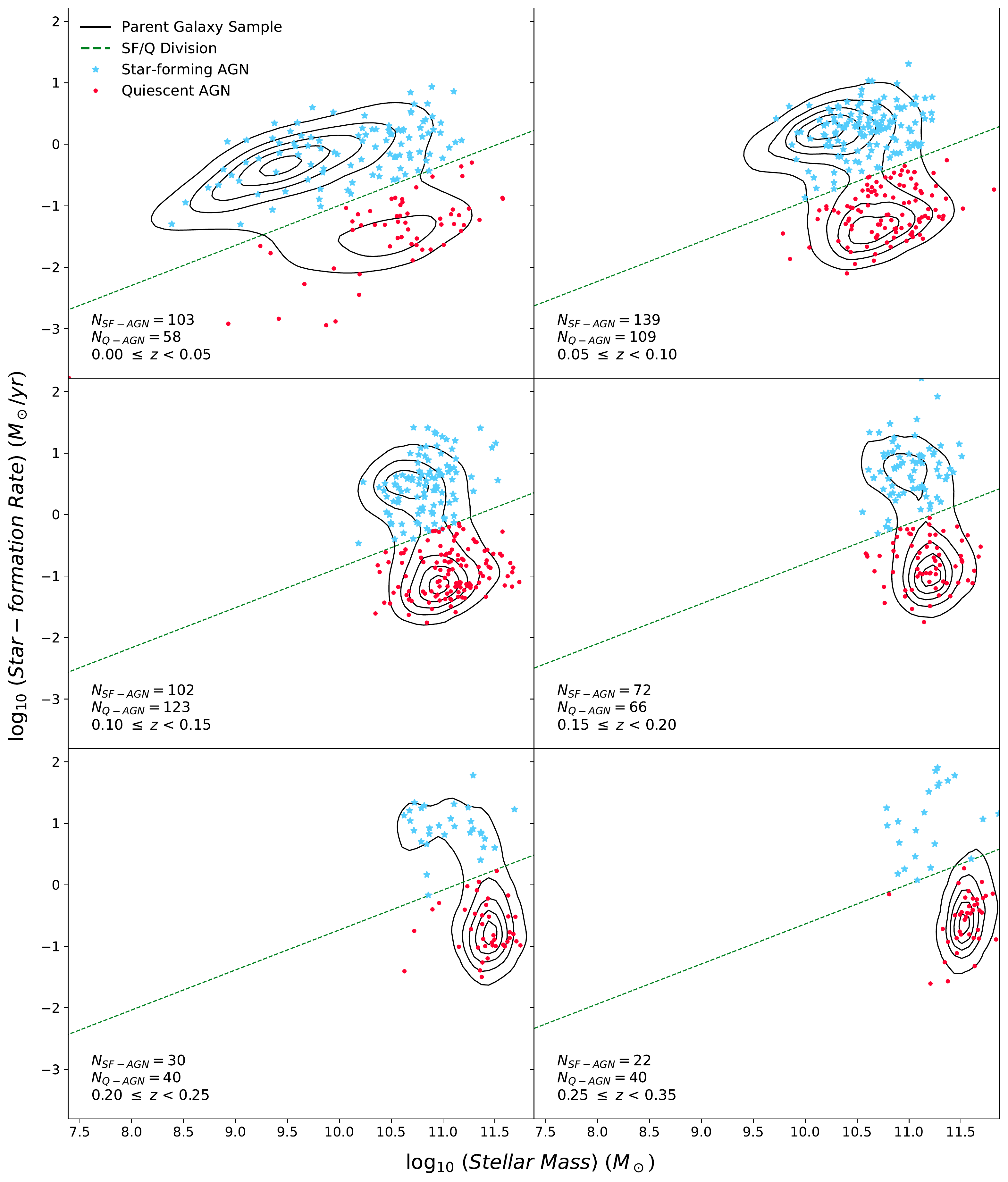}

        \caption[Definition of star-forming and quiescent galaxy populations]{The AGN sample contrasted with the underlying galaxy population (black contours, containing 90\%, 70\%, 50\%, 30\%, and 10\% of the population). The green dashed line describes the changing minimum of the bimodal SFR distribution for this galaxy population and is used to determine whether an AGN is "star-forming" (blue stars) or "quiescent" (red circles). See section \protect \ref{subsec:SF_or_Q} for more information on how this was calculated.}
	\label{fig:SF_Classification}
\end{figure*}

Figure \ref{fig:SF_Classification} shows the results of this analysis. Our AGN sample is plotted as large, coloured points on the mass and SFR plane and contrasted with the underlying MPA-JHU galaxy population. Each panel contains a green line, described by equation \eqref{eq:SF_Split}, which is used to split these AGN into star-forming (blue stars) and quiescent (red circles). Any objects that lie above the line are classified as star-forming, those below it are quiescent. 

\subsection{SFR relative to the Main Sequence}
\label{subsec:SFR_rel_MS}
Dividing the sample into star-forming and quiescent galaxies is useful to compare the general effect of star-formation on AGN activity. However, to understand this effect in greater detail we extended the above analysis to further divide the galaxy sample by its changing level of star-formation. For this quantity we shifted equation \eqref{eq:SF_Split} up to the galactic main sequence of star formation at a given redshift \citep[See][]{Aird19}. It has the form,
\begin{equation}
    \log_{10} (\mathrm{SFR_{MS}}) = \log_{10}(\mathrm{SFR}) - 0.65(\log_{10} M_\mathrm{*} - 10) + \frac{z - 0.05}{0.77}
    \label{eq:MS_Defn}
\end{equation}

\begin{figure}
	\centering
     	\includegraphics[width=\columnwidth]{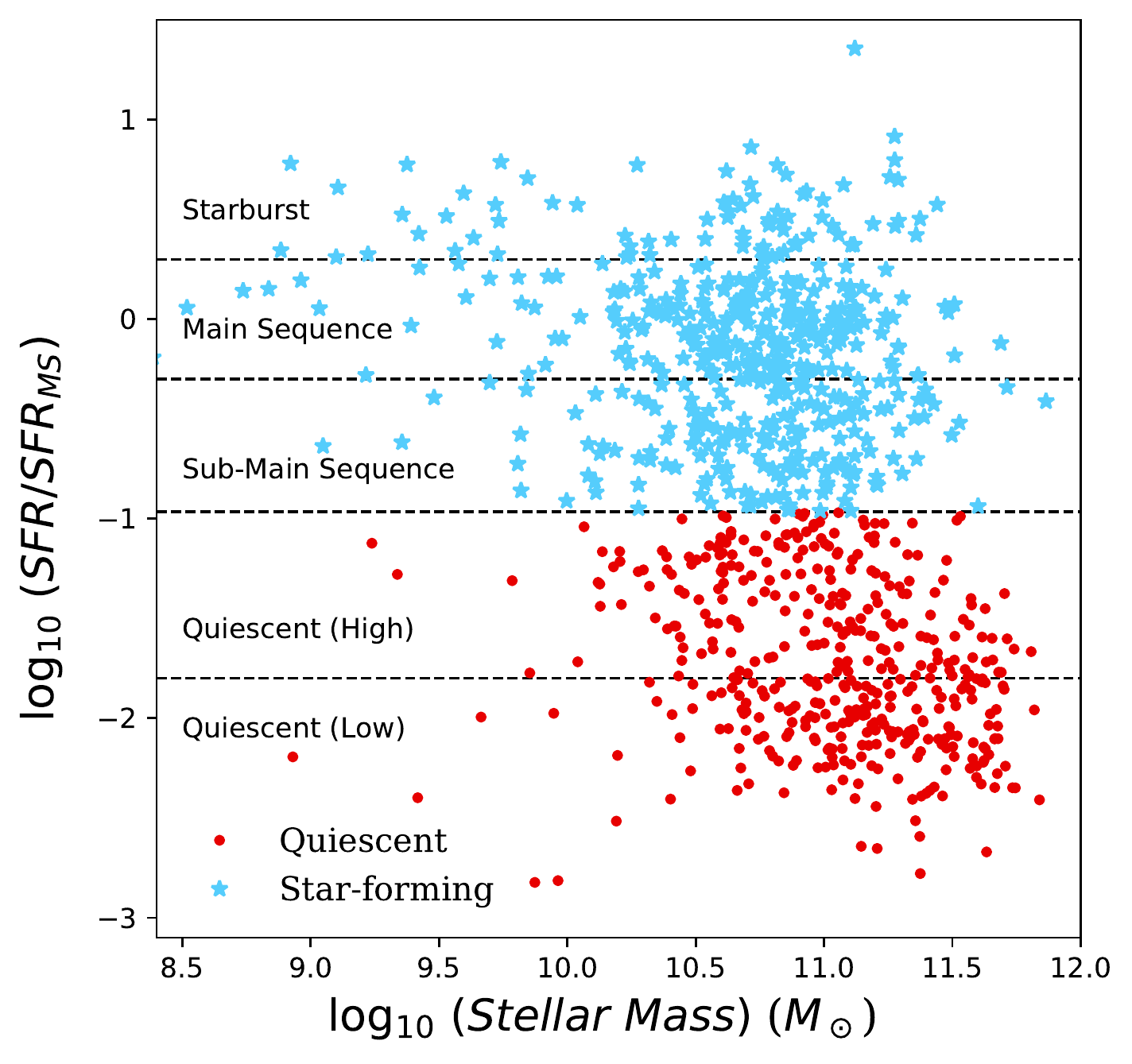}

		\caption[Definitions of $\log_{10} (\mathrm{SFR/SFR_{MS}})$]{Defining the of $\log_{10} (\mathrm{SFR/SFR_{MS}})$ groups for the AGN sample. Overlaid are black, dashed lines describing bins of $\log_{10}(\mathrm{SFR/SFR_{MS}})$ calculated based on the underlying galaxy population. Star-forming AGN are shown as blue stars, and quiescent AGN are red circles. Section \ref{subsec:SFR_rel_MS} provides an explanation of how these limits were chosen. }
	\label{fig:SF_Defns}
\end{figure}

With this equation we could establish 5 bins of $\log_{10} (\mathrm{SFR/SFR_{MS}})$ to track the changing level of star-formation in the sample:
 
 \begin{itemize}
     \item \textbf{Starburst}: Star-forming galaxies with excess star-formation relative to the main sequence ($\log_{10}(\mathrm{SFR/SFR_{MS}}) > 0.3$) \linebreak
     
     \item \textbf{Main Sequence}: Star-forming galaxies with SFRs consistent with the main sequence ($-0.3 \leq \log_{10}(\mathrm{SFR/SFR_{MS}}) \leq 0.3$) corresponding to 50\% of the total star-forming galaxy population \linebreak
     
     \item \textbf{Sub-Main Sequence}: Galaxies with SFRs lower than the bulk of the main sequence ($-0.965 \leq \log_{10}(\mathrm{SFR/SFR_{MS}}) < -0.3$) consisting of weak star-forming galaxies \linebreak
     
     \item \textbf{Quiescent (High)}: Quiescent galaxies with SFRs in the top 50\% of this population ($-1.8 \leq \log_{10}(\mathrm{SFR/SFR_{MS}}) < -0.965$) \linebreak
     
     \item \textbf{Quiescent (Low)}: Quiescent galaxies with SFRs in weakest 50\% of that population ($\log_{10}(\mathrm{SFR/SFR_{MS}}) < -1.8$)

\end{itemize}
Figure \ref{fig:SF_Defns} shows how each bin maps onto the AGN sample. By taking this approach we have ensured there are sufficient numbers of observed AGN in each bin and that, for a fixed redshift, an increase in the $\log_{10}(\mathrm{SFR/SFR_{MS}})$ tracks only the effects of SFR.

\section{BPT Classification}
\label{sec:BPT}
AGN activity impacts the host galaxy's emission across the electromagnetic spectrum. The BPT diagnostic \citep{BaldwinPhillipsTerlevich81} is a commonly used technique which can identify the dominant source of ionising radiation in the optical part of the spectrum. It compares the ratios of various emission lines to determine whether star formation, AGN or a composite of both processes are the likely dominant source of ionisation in any given galaxy. \cite{Birchall20} found that this diagnostic missed around 85\% of our X-ray selected AGN in dwarf galaxies. \cite{Birchall22} found that BPT selection missed a similar proportion of AGN in low mass galaxies but that its accuracy increased towards higher stellar masses. In this section, we investigate how star formation affects the accuracy of this diagnostic.\\ 

\begin{figure} 
\includegraphics[width=\columnwidth]{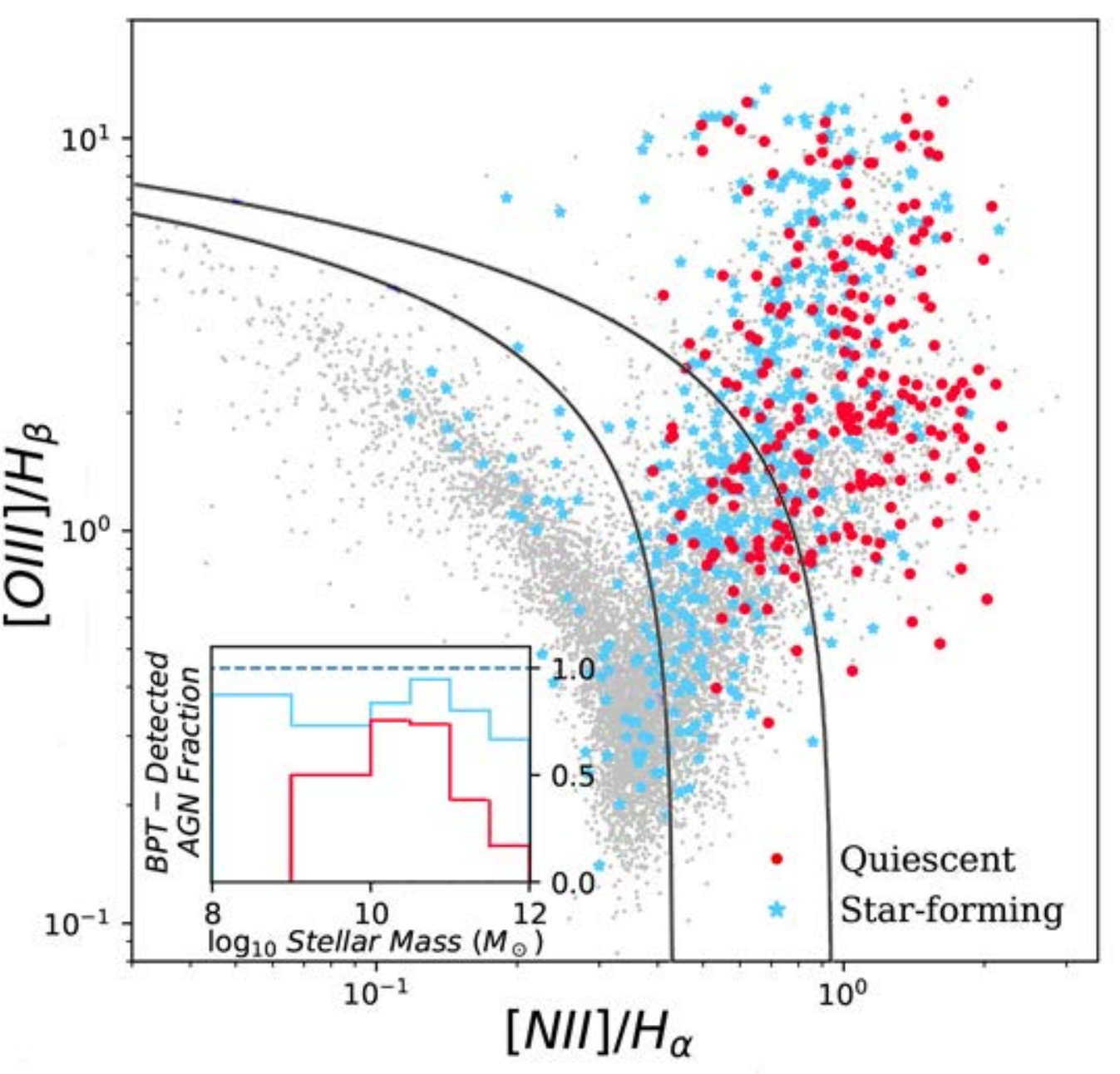}
\caption{BPT diagram for the 658 AGN with significant detections in all four emission lines overlaid onto the underlying galaxy population in grey. Star-forming AGN are shown as blue stars, and quiescent AGN are red circles. The inset figure shows the fraction of AGN that have sufficiently strong emission lines to be detected by the BPT diagnostic, split up star-forming category and as a function of stellar mass. Black lines separate the AGN hosts into different classifications: star-formation-dominated sources lie in the bottom-left, AGN-dominated sources in the top-right, and those with composite spectra are in the central region \citep{Kewley01, Kauffmann03b}.}
\label{fig:BPT}
\end{figure}

Of the 917 AGN hosts we identified in section \ref{sec:data}, 658 had strong detections ($\frac{\mathrm{Line\ Flux}}{\mathrm{Line\ Flux\ Error}} > 3$) in the required emission lines. These were used in our BPT analysis. In figure \ref{fig:BPT}, we show the BPT and X-ray-selected AGN on the BPT plane, with star-forming AGN as blue stars, and their quiescent counterparts as red circles. Black lines separate the AGN hosts into different classifications: objects with ionisation signatures predominately from star formation lie in the bottom-left, those dominated by AGN emission lie in the top-right, and those with composite spectra are in the central region \citep{Kewley01, Kauffmann03b}. A subset of the MPA-JHU galaxies are shown underneath these points, in grey, to illustrate the underlying BPT distribution.

The vast majority of the X-ray selected star-forming AGN have significant enough detections in all the BPT emission lines to make it the BPT diagram. This population can be clearly seen across the full extent of the diagnostic. However, under half of the X-ray selected quiescent AGN have significant enough detections to make it onto the BPT diagram. This population is concentrated largely in the AGN region.\\

By requiring a significant detection in each BPT emission line, objects with relatively large amounts of optical ionisation are much more likely to make it onto figure \ref{fig:BPT}, regardless of its source. However, this is not reflected in the X-ray luminosities. There is no systematic difference in the X-ray luminosity distributions of BPT-detected and non-detected sources for either the full AGN population or just the quiescent AGN.  \\

By definition, star-forming AGN produce significant amounts of optical emission from both forming stars and AGN activity. Thus, these objects will have higher levels of optical emission on average. This results in a higher fraction of these objects having significant enough emission lines to make it onto the BPT diagram. It also means that these objects can appear across all classifications, although, a BPT classification of 'composite' or 'star-formation-dominated' does not exclude the presence of AGN activity \citep{Birchall20, Birchall22}.\\

Quiescent galaxies, on the other hand, are defined by their reduced levels of star formation. This means that the only route by which they can appear on the BPT diagram is through significant AGN activity. This explains why these objects are concentrated in the AGN region. It could also explain why these objects are disproportionately missing from the BPT diagnostic: if there is little star formation and the AGN activity is optically quiet, then there is no source to produce strong enough emission lines to make it onto the BPT diagnostic. \\

To investigate this, we first analysed the individual emission lines and found that H$\beta$ most frequently missed the significance threshold. \cite{CidFernandes10} also found a significant population of SDSS galaxies were missed out from the BPT diagram due to lack of significant H$\beta$ line detections. In addition, it was galaxies that would have been classified as AGN (should the significance criterion not have been required) that were most affected. \\

We then looked at how the fraction of BPT-detected AGN changes with stellar mass and star-forming classification. The results are inset in figure \ref{fig:BPT}. The blue, star-forming distribution shows high proportions of BPT-detected AGN, reaching $94\%$ in the $10.5 \leq \log_{10}\ (M_\mathrm{*}/M_\mathrm{\odot}) \leq 11$ bin and never dropping below $67\%$. However, the red, quiescent distribution peaks at 77\% in the $10 \leq \log_{10}\ (M_\mathrm{*}/M_\mathrm{\odot}) \leq 11$, before rapidly dropping to $17\%$ in the highest mass bin. There was only one quiescent AGN detected in the lowest mass bin which was missed, hence the $0\%$ detection rate. \\

Since, by definition, quiescent AGN are not forming as many new stars as their star-forming counterparts, stellar absorption lines, like H$\beta$, could dominate their spectra. These absorption lines will be most pronounced in the spectra of the highest mass galaxies as they contain the largest amount of stellar material. Thus, weak AGN emission lines could be dominated by strong absorption lines and produce the significant drop in the fraction of high mass, quiescent AGN shown on the BPT diagram in figure \ref{fig:BPT}. This suggests that the BPT diagnostic is not effective at identifying optically-weak emission from AGN, especially in quiescent galaxies.

\section{Specific Black Hole Accretion Rate}
\label{sec:sBHAR}
 AGN are powered by the accretion of matter onto a central supermassive black hole. Measuring the accretion rate is important as observing only the absolute X-ray luminosity can provide a biased picture, especially when examining the AGN content of galaxies that span a broad mass range. Consider two black holes, one black hole is growing at a higher accretion rate in a lower mass galaxy, another black hole with a lower accretion rate in a high mass galaxy. It is possible that these two galaxies could emit the same X-ray luminosity, obscuring the activity occurring at the central black hole. Thus, to break this degeneracy, we need to investigate the specific black hole accretion rate (sBHAR), $\lambda_\mathrm{sBHAR}$. This quantity compares the  bolometric AGN luminosity with an estimate of the black hole's Eddington luminosity to provide an estimate of how efficiently material is being accreted. It has the following form,
\begin{equation} \label{eq:sBHAR}
    \lambda_{\mathrm{sBHAR}} = \frac{25 L_{\mathrm{2-10 keV}}}{1.26 \times 10^{38} \times 0.002M_*} \approx \frac{L_{\mathrm{bol}}}{L_{\mathrm{Edd}}}
\end{equation}

and is taken from \cite{Aird12}. Using $0.002 M_*$ implies a perfectly constant correlation between the masses of the SMBH and stellar bulge which, in reality, is dependent on galaxy morphology, among other properties \citep[e.g.][]{BlantonMoustakas09,Jahnke09}. However, our intention when using this correlation is to present an Eddington-scaled accretion rate quantity rather than accurately recreate an Eddington ratio. By using this scaling to calculate sBHAR - a tracer of the rate of black hole growth relative to the host galaxy's stellar mass - it allows for ease of comparison with the literature.\\

\begin{figure*}
    \centering
    \includegraphics[width=.6\paperwidth]{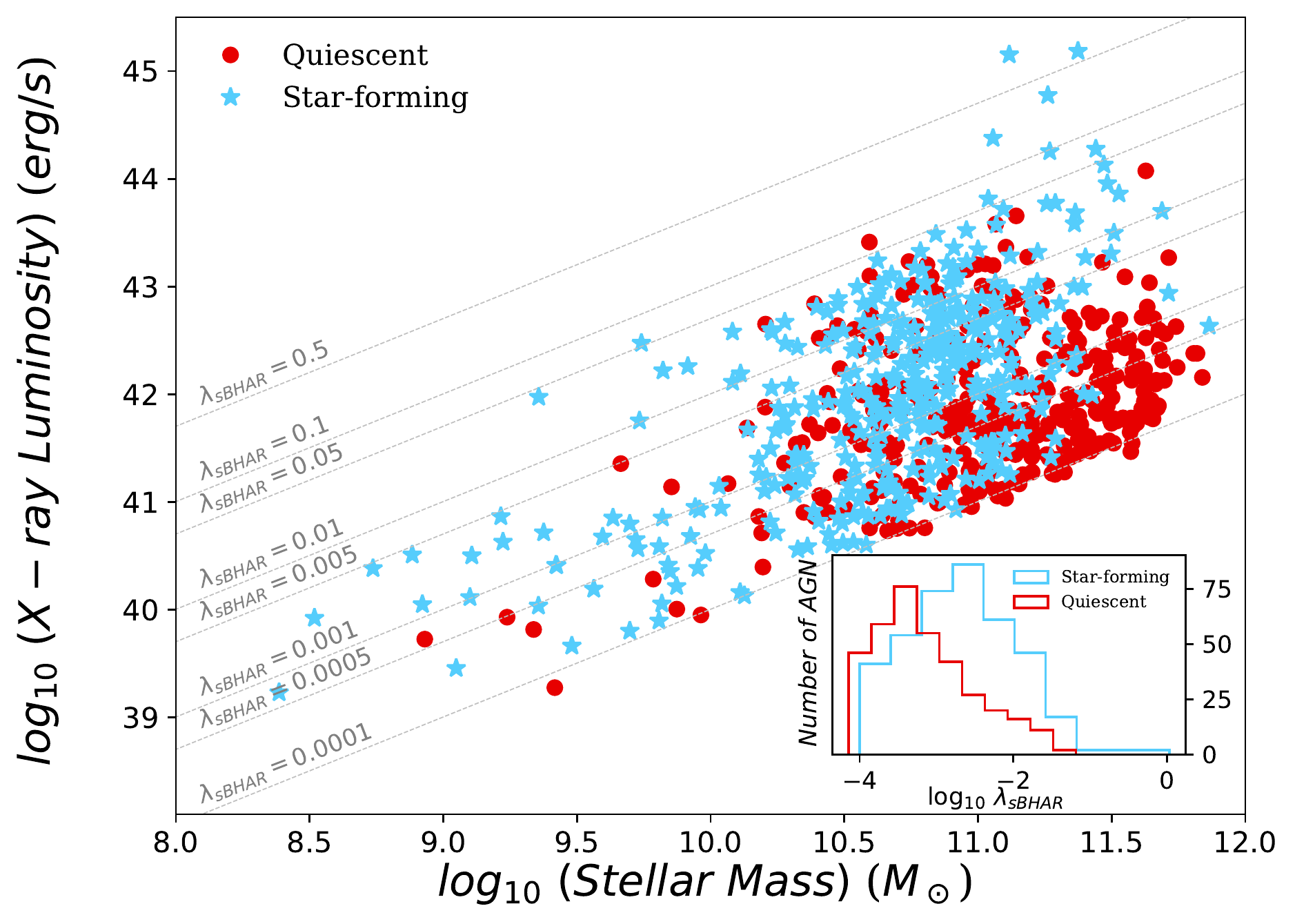}
    \caption{Stellar mass against the observed X-ray luminosity for the 917 AGN detected in section \ref{sec:data}. Star-forming AGN are shown as blue stars, and quiescent AGN are red circles. Several grey lines of constant sBHAR have also been plotted for reference. The inset panel shows how the AGN population is distributed along the sBHAR axis when split by star-forming classification. This distribution has not been corrected to account for X-ray sensitivity limits, for more information on this see section \ref{sec:Prob_Dists}.}
    \label{fig:sBHAR}
\end{figure*}

Figure \ref{fig:sBHAR} shows the observed AGN sample distributed on a stellar mass and X-ray luminosity plane, and coloured based on their star-forming classification. Star-forming AGN are shown as blue stars, and quiescent AGN are red circles. Lines of constant sBHAR are shown to indicate its approximate value for different masses and luminosities. The clearest difference between star-forming and quiescent AGN is in the highest mass galaxies ($\log_{10} \mathrm{M_*/M_\odot} \gtrsim 11$). Here the brightest, most actively accreting AGN are in star-forming galaxies, and the dimmer, weakly accreting AGN are in quiescent hosts. This effect is repeated across the stellar mass range to a lesser degree, with star-forming AGN generally having higher values of sBHAR. This is evident in the inset panel, where we plotted the sBHAR distributions split by star-forming classification. It is clear that quiescent AGN are dominated by relatively low accretion rates, whereas star-forming AGN are much more evenly spread and peak at much higher rates. In fact, for quiescent AGN, the mean $\log_{10} \lambda_\mathrm{sBHAR} \sim -3.1$, whereas for  star-forming AGN, the mean $\log_{10} \lambda_\mathrm{sBHAR} \sim -2.7$. Whilst this sample is subject to a selection bias, it clearly shows there is an increased level of accretion activity in star-forming galaxies. This result has been seen in numerous samples at similar \citep[][]{Torbaniuk2021} and higher redshifts \citep[e.g.][]{Chen13, Yang17}. 


\section{Completeness-corrected Probability Distributions}
\label{sec:Prob_Dists}
No AGN selection technique is perfect. Our approach, briefly described in section \ref{sec:data}, aimed to overcome the preference that a flat luminosity threshold has towards selecting higher mass galaxies (which typically have higher X-ray luminosities) by modifying the threshold value based on the host galaxy properties. However, figure \ref{fig:sBHAR} shows that despite our efforts the sample is still dominated by higher mass galaxies.  
\cite{Birchall20} outlines, in detail, the method we developed to reduce the effects of this bias. By correcting for 3XMM's varying detection sensitivity, this method allows us to investigate how the underlying distribution of AGN varies with changes in host galaxy properties. \cite{Birchall22} showed that the X-ray luminosity probability distributions were still subject to an apparent observational bias, despite the completeness corrections applied. So, we only consider the sBHAR-based probability distributions. In this section, we will briefly describe how the completeness corrections are calculated, how they are applied to the observational data to create these distributions, and what they tell us about the effect of star formation on the incidence of AGN in the nearby Universe. See \cite{Birchall20, Birchall22} for a more comprehensive description of this process.

\subsection{Creating the Probability Distributions}
\label{sub_sec:prob_dist_create}
The detection sensitivity of 3XMM varies significantly across its observed sky area. This variation brings the possibility that lower luminosity AGN in this region may have been missed because 3XMM is insufficiently sensitive to detect them. To correct for this, we used Flix, 3XMM's upper limit service, to measure the $0.2 - 12\ \mathrm{keV}$ upper limit at the centres of the 22,079 MPA-JHU galaxies in the 3XMM footprint. We restricted detections to the PN8 band which reduced our AGN sample from 917 to 739. Using the Flix upper limit and the MPA-JHU redshift and stellar mass, we calculated the X-ray luminosity upper limit and $\lambda_\mathrm{sBHAR}$ for each galaxy. With these properties, we could construct a cumulative histogram normalised by the total number of galaxies in the current mass and redshift range. These luminosity sensitivity functions allowed us to determine the fraction of galaxies where an AGN accreting above a given rate could be detected. \\

\begin{figure*}
    \centering
    \includegraphics[width=.8\paperwidth]{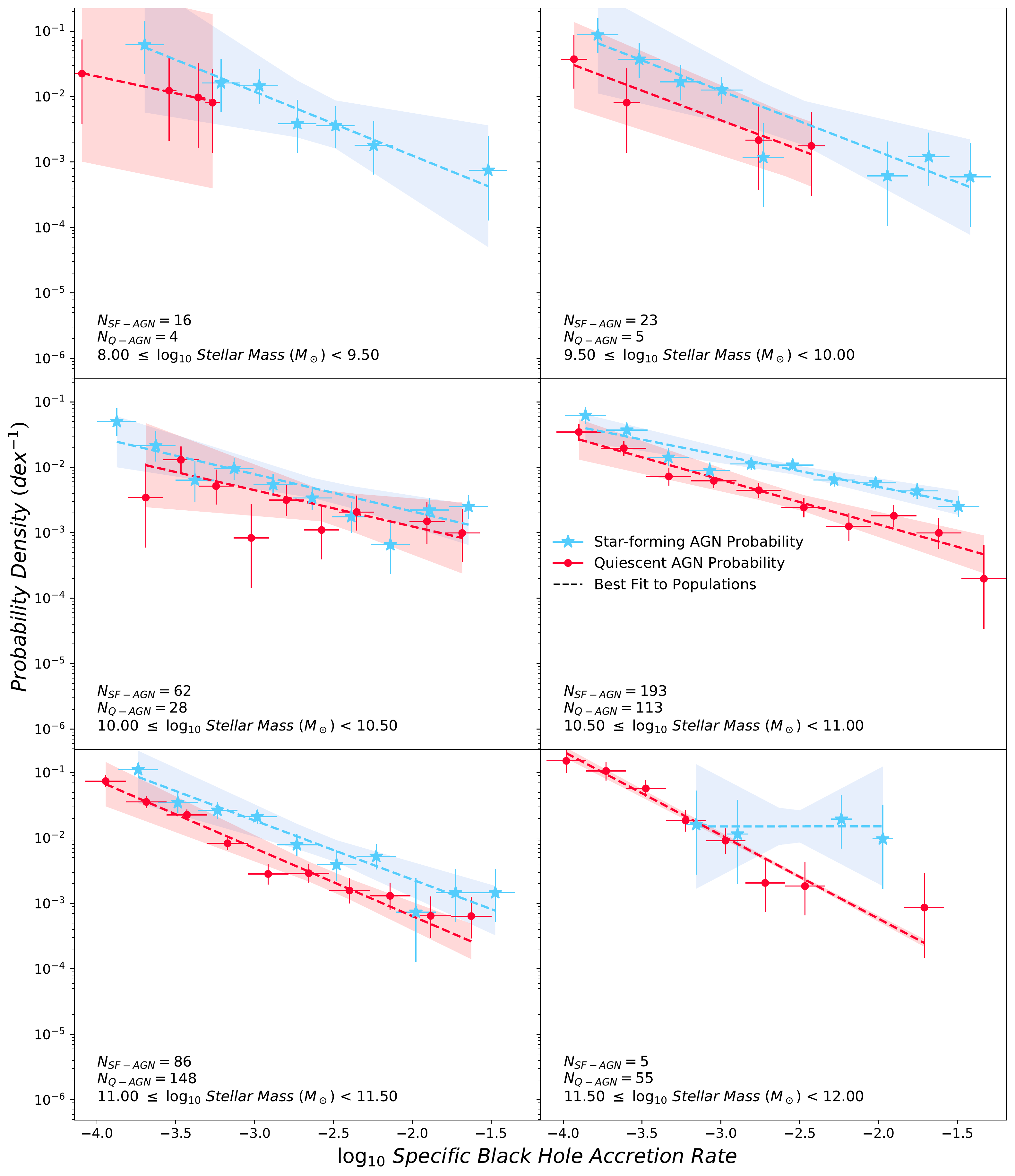}
    \caption{Probability of finding an AGN in the nearby Universe, using the completeness-corrected AGN samples, as a function of $\mathrm{\lambda_{sBHAR}}$ and stellar mass. The star-forming AGN population are shown as blue stars, and the quiescent AGN population are red circles. Power laws (dashed lines) have been fit to the data in each panel and displayed alongside their $1\sigma$ uncertainty (pale region). Information about the number of PN-8-detected AGN and parent galaxies in the each mass range are printed on the panels. See section \ref{sub_sec:prob_dist_create} for more details on how these plots were constructed. Table \ref{tab:Prob_Dist_Coeffs} lists the coefficients required for equation \eqref{eq:powerlaw} to recreate these distributions.}
    \label{fig:Example_Prob_Dist}
\end{figure*}

Using these sensitivity functions, we can, for example, calculate the probability of finding an AGN in a given galaxy as a function of specific black hole accretion rate, in bins of increasing stellar mass. To construct this distribution, we first split up the AGN sample into a series of stellar mass bins. These are then further broken as a function of $\lambda_\mathrm{sBHAR}$. This gives us an observed AGN count distribution. We also created a bespoke sensitivity function for each of these stellar mass bins. From this, we constructed a binned probability distribution by dividing the number of AGN in a given galaxy sub-sample and sBHAR bin, $N_{\mathrm{AGN},\ i}$, by the number of galaxies where such an sBHAR would be detectable, $N_\mathrm{gal}$. Thus,
\begin{equation}
   p(\lambda_{\mathrm{sBHAR},\ i}) = \frac{N_{\mathrm{AGN},\ i}}{N_\mathrm{gal} (\lambda_{\mathrm{sBHAR},\ lim} > \lambda_{\mathrm{sBHAR},\ i})}
\end{equation}

where $p(\lambda_{\mathrm{sBHAR},\ i})$ is the probability of observing an AGN in a given galaxy sub-sample, accreting above the limiting accretion rate $\lambda_{\mathrm{sBHAR},\ lim}$. The results of this process are shown in figure \ref{fig:Example_Prob_Dist}, the probability of finding an AGN as a function of sBHAR in bins of increasing mass. Reference information is also printed on each panel including the size of each AGN population and the stellar mass range of the galaxies included therein. \\

By applying correction fractions, extracted from the $\lambda_\mathrm{sBHAR}$ sensitivity function, to the observed AGN counts we were able to provide robust measurements of the true incidence of AGN within the nearby galaxy population. Figure \ref{fig:Example_Prob_Dist} shows the results of this calculation. It is clear that AGN exist across the stellar mass range despite the favouring of higher mass AGN seen in figure \ref{fig:sBHAR}. In addition, we see that AGN populations are well described by power law distributions, as found in \cite{Birchall20} and \cite{Birchall22}, with AGN being found predominantly at lower $\lambda_\mathrm{sBHAR}$ \citep[see also][and others]{Aird12}. \\

Errors on the data points are found using the confidence limits equations presented in \cite{Gehrels86} so the size of the error is proportional to the number of detected AGN in a given bin.\\

Power laws were fit to the data in each of these panels with the following form,
\begin{equation}
    \label{eq:powerlaw}
      p(\lambda) = A \left( \frac{\lambda}{\lambda'} \right) ^k d \log_{10} \lambda
\end{equation}

where $p(\lambda)$ is the probability of observing an AGN as a function of $\lambda_\mathrm{sBHAR}$ and centred on a value $\lambda’$. Each power law is centred on the median accretion rate for the sample, $\log_{10} \lambda' = -2.55$. The star-forming AGN population are shown as blue stars, and the quiescent AGN population are red circles The power laws are shown as dashed lines, indicating how the probability of finding an AGN changes as a function of $\lambda_\mathrm{sBHAR}$. The error regions surrounds each power law and was calculated by performing a $\chi^2$ fit with equation \eqref{eq:powerlaw} to the data points and errors in each bin. Fit parameter errors were estimated by taking the square-root of the covariance matrix’s diagonal. With this we could outline the extent of the uncertainty in each fit. Encouragingly, the power law provides a good fit in nearly all of the bins. 

\subsection{Probability Distribution Comparison}
\label{sec:PD_comp}
Using the method described above we have created a series of probability distributions investigating how other host galaxy properties might affect the incidence of AGN in the nearby Universe. Figure \ref{fig:Prob_Dist_Comparison} adds probability distributions for redshift and $\log_{10} \mathrm{(SFR/SFR_{MS})}$, split up into quiescent (left-hand column) and star-forming (right-hand column) AGN.  Each panel contains numerous probability distributions, represented by dashed lines, which are coloured to indicate the respective stellar mass (top row), redshift (middle row) or  $\log_{10}(\mathrm{SFR/SFR_{MS}})$ (bottom row) bin. The best-fit coefficients and associated errors used to calculate these distributions are presented in appendix \ref{tab:Prob_Dist_Coeffs}. The error regions were all fit to data points but were not included on figure \ref{fig:Prob_Dist_Comparison} to aid clarity. The redshift and $\log_{10} \mathrm{(SFR/SFR_{MS})}$ distributions with data points are presented in appendix \ref{app:Prob_Dists}. 

\subsubsection{Normalisation}
\label{subsec:PD_SF_Norm}
The normalisation of the distributions in figure \ref{fig:Prob_Dist_Comparison} do not change significantly. There is some evidence of increase in the overall probability of finding AGN in star-forming galaxies as we shift towards higher redshifts and star formation rates, but the magnitude of this effect appears to be fairly small. \\

Not all of these distributions cover the full extent of $\log_{10} \lambda_\mathrm{sBHAR}$ axis. This can be seen most clearly in the stellar mass row, with the distributions describing the lowest mass, quiescent AGN and highest mass, star-forming AGN. In addition, the star-forming column of both the redshift and $\log_{10} \mathrm{(SFR/SFR_{MS})}$ rows show a slight shift towards higher accretion rates at higher values of the respective property. These changes are due to the limited number of observations in these bins. However, this does highlight changes in galaxy population across the star-forming classifications. In the stellar mass row, we see that star-forming galaxies dominate at lower stellar masses and quiescent at higher masses. And in the redshift and $\log_{10} \mathrm{(SFR/SFR_{MS})}$ rows, a slight favouring of higher accretion rate galaxies. 

\subsubsection{Slope}
\label{subsec:PD_SF_Slope}
\begin{figure*}
    \centering
    \includegraphics[width=.73\paperwidth]{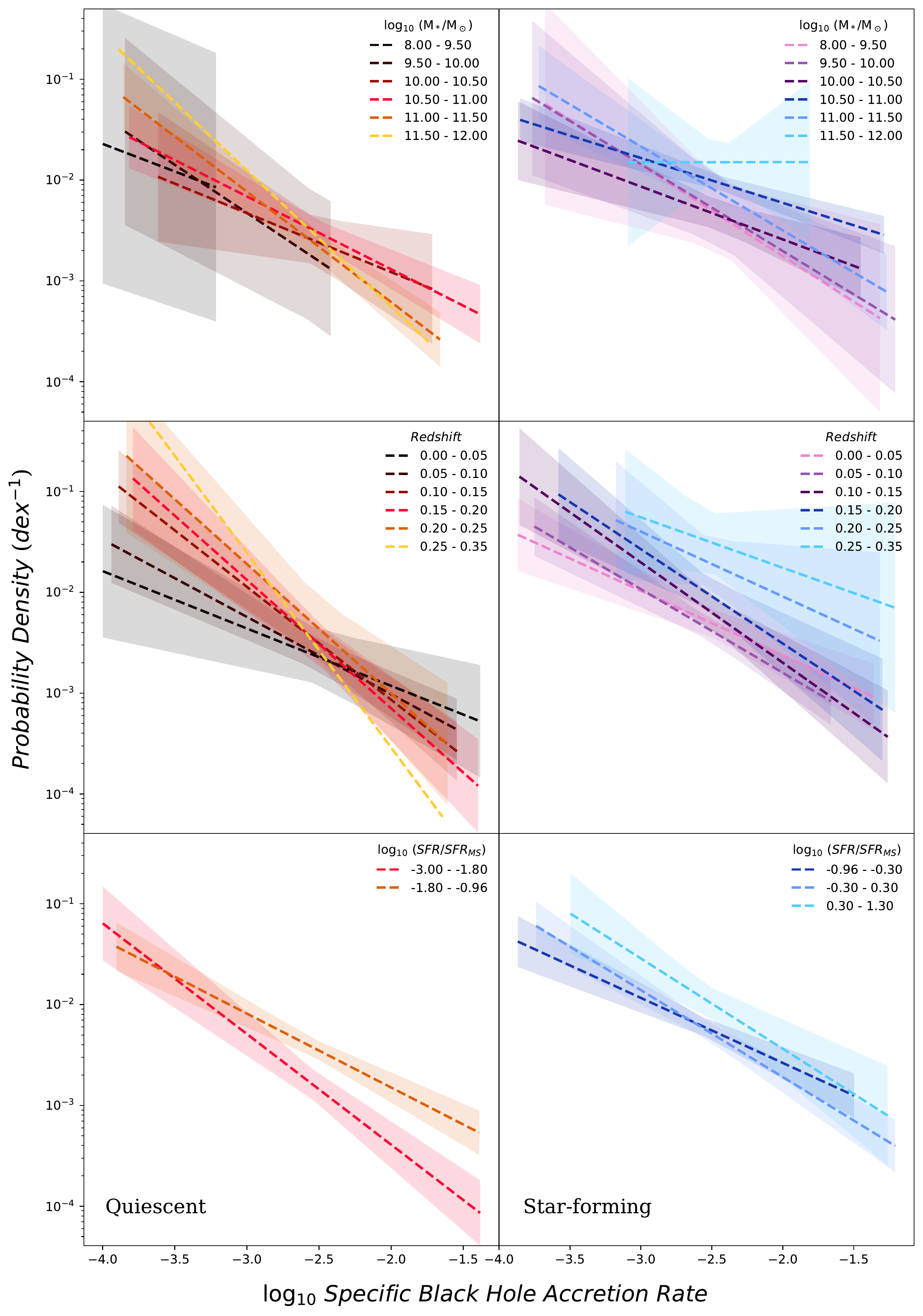}
    \caption{Comparison of the full range of sBHAR-dependent probability distributions. The left-hand column plots the distributions of quiescent AGN, the right-hand column plots the distributions of the star-forming AGN. The distributions are binned as a function of stellar mass (top row), redshift (middle row) and  $\log_{10} (\mathrm{SFR/SFR_{MS}})$ (bottom row). See section \ref{sub_sec:prob_dist_create} for more details on how these plots were constructed. Table \ref{tab:Prob_Dist_Coeffs} lists the coefficients required for equation \eqref{eq:powerlaw} to recreate these distributions.}
    \label{fig:Prob_Dist_Comparison}
\end{figure*}

The clearest changes in slope can be seen in the redshift row of figure \ref{fig:Prob_Dist_Comparison}. The quiescent distributions are much steeper than those composed of star-forming AGN. There is also a distinct steepening between the quiescent distributions, showing that lower accretion rates are favoured with increasing redshift. The star-forming distributions, however, do not show this trend. In fact, the highest redshift, star-forming distributions appear significantly flatter than their quiescent counterparts, suggesting that star formation facilitates higher accretion rates at higher redshifts. \\

There is not much change in slope in the $\log_{10} (\mathrm{SFR/SFR_{MS}})$ row. Given the nature of this quantity the quiescent/star-forming division occurs along the $\log_{10} (\mathrm{SFR/SFR_{MS}})$ axis, so there are only two bins in the left panel and three on the right. \\

There does not appear to be any consistent trend within the stellar mass row, nor any clear effect when comparing the quiescent and star-forming populations. Whilst the highest mass, star-forming bin has a distinctly flat gradient, it is poorly constrained and is consistent with the lower mass bin within the errors.  

\section{AGN Fractions}
\label{sec:fractions}
\begin{figure*}
	\centering
     	\includegraphics[width=.6\paperwidth]{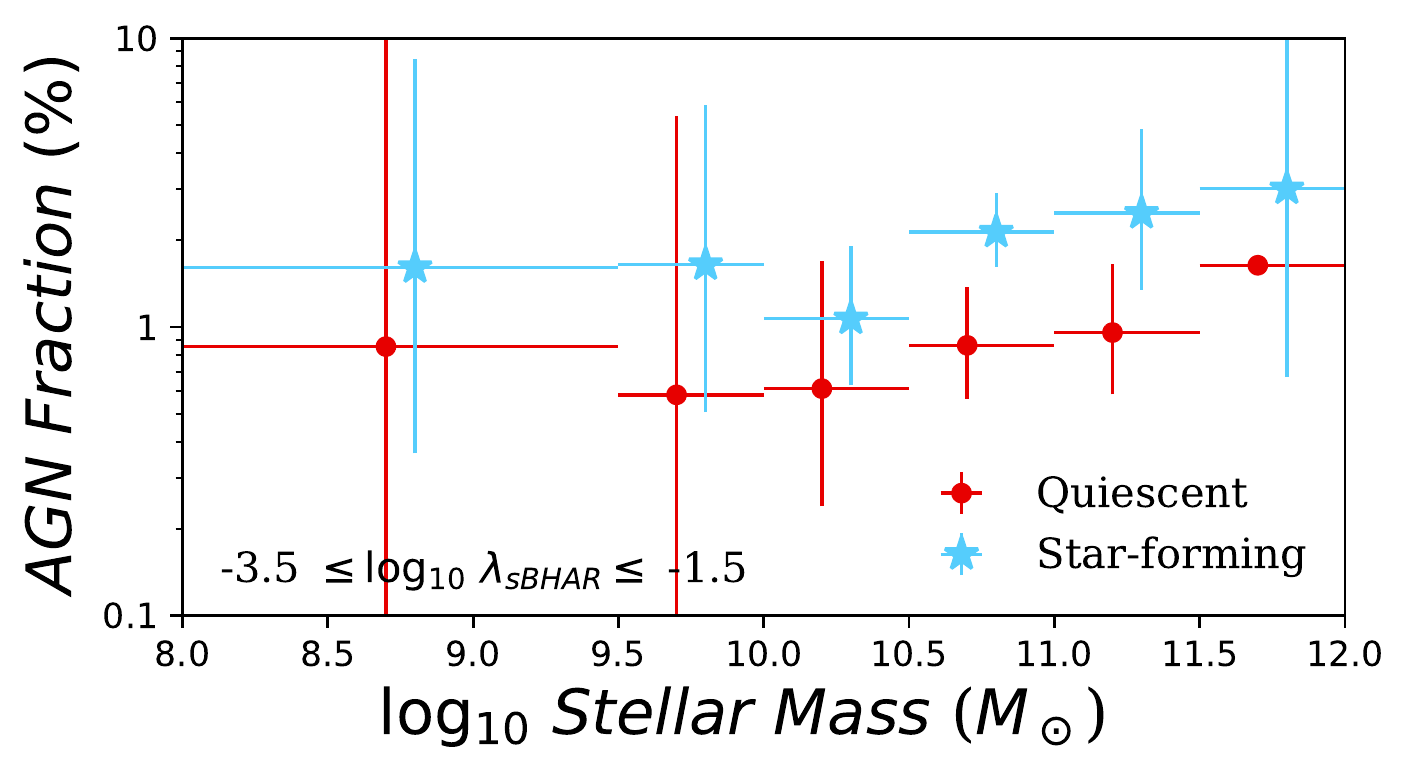}
     	
     	\includegraphics[width=.6\paperwidth]{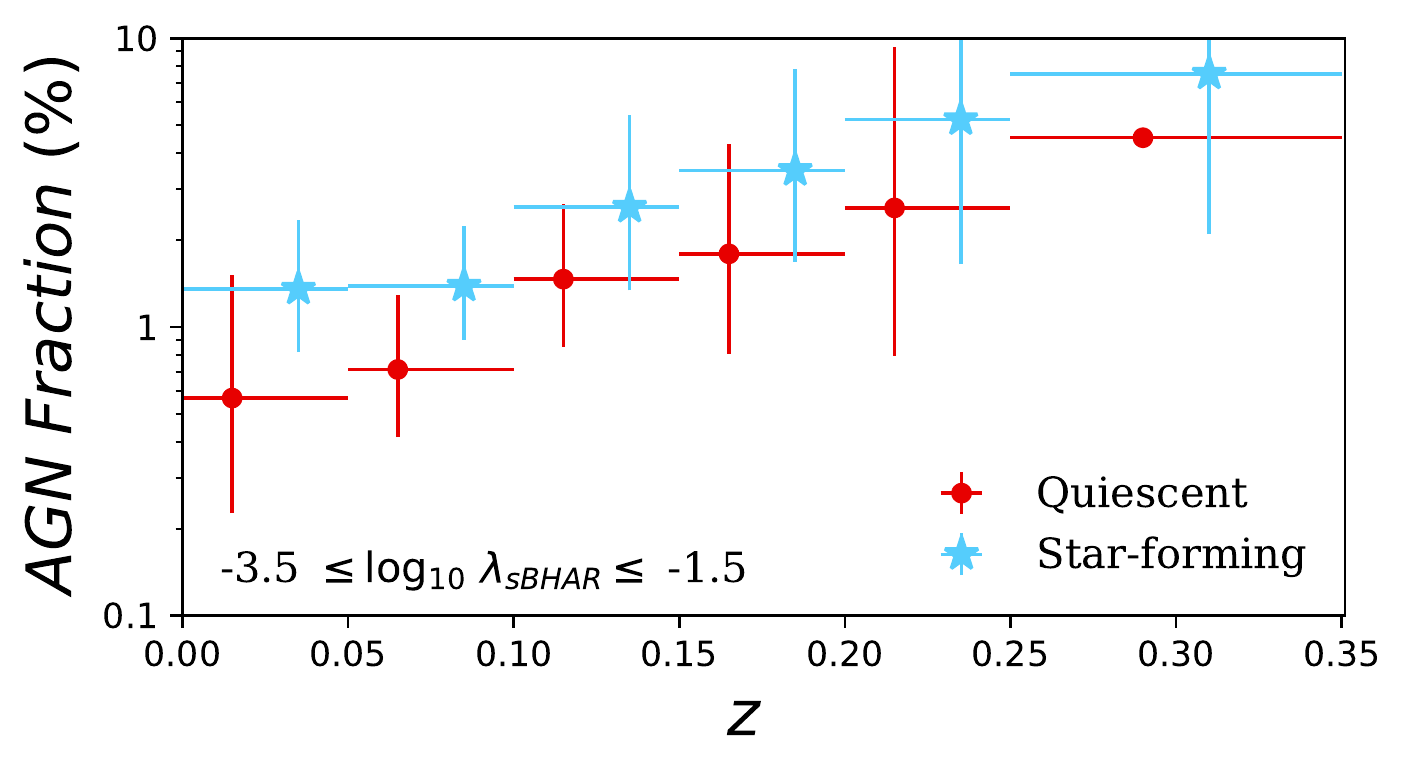}
     	
     	\includegraphics[width=.6\paperwidth]{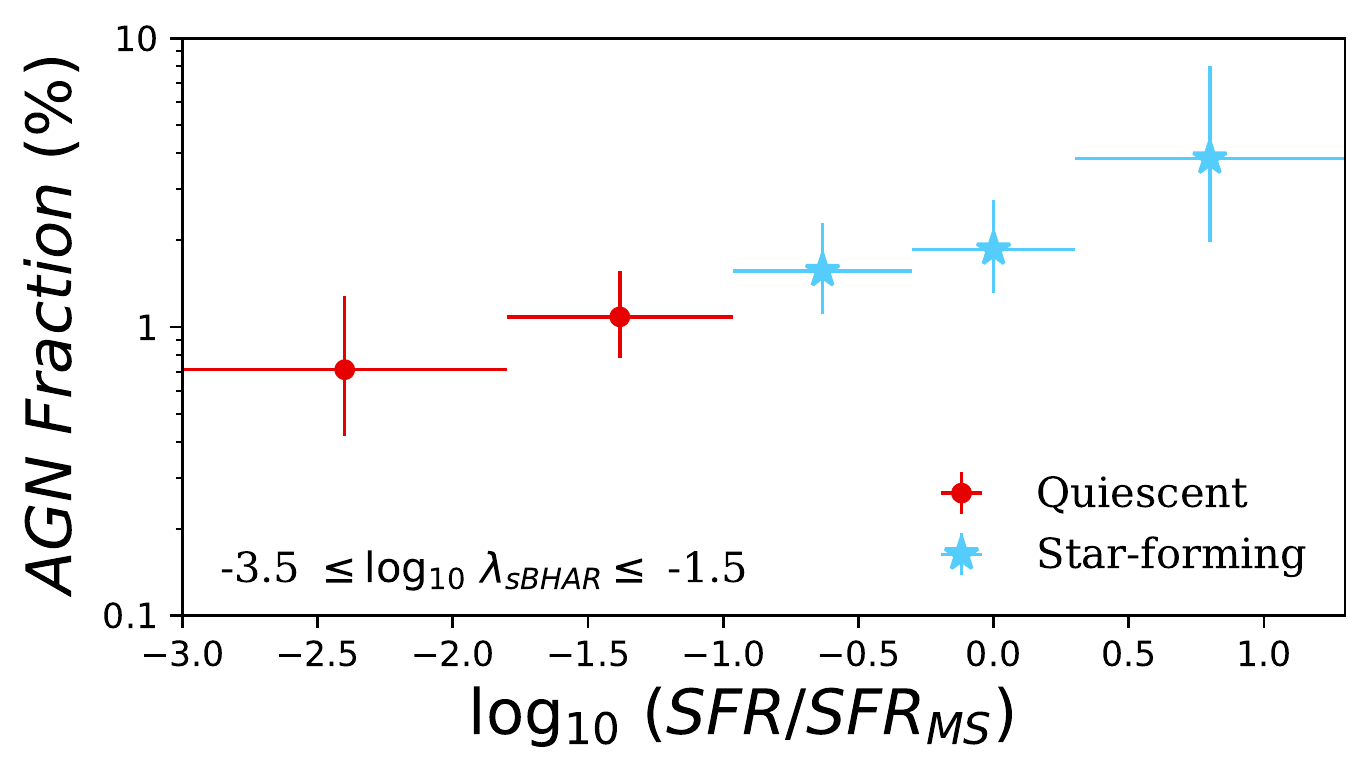}
     	
		\caption[MPA-JHU AGN fraction as a function of stellar mass, redshift \& $\log_{10} (\mathrm{SFR/SFR_{MS}})$, split by star-forming classification]{AGN fraction as a function of stellar mass (top), redshift (middle) and $\log_{10} (\mathrm{SFR/SFR_{MS}})$ (bottom). In each panel star-forming AGN fractions are shown as blue stars, and quiescent AGN fractions are red circles. The sBHAR ($\lambda_\mathrm{sBHAR}$) integration limits are also displayed in the bottom-left corner.}
	\label{fig:Overall_Frac}
\end{figure*}

Integrating under the probability distributions shown in figure \ref{fig:Prob_Dist_Comparison}, allowed us to further analyse how star formation affects the AGN population in the nearby Universe. Figure \ref{fig:Overall_Frac} shows the AGN fraction as function of each property, split up into star-forming and quiescent populations. As before we are only considering the sBHAR-derived fractions; the corresponding integration limits are shown in the bottom-left corner of each panel. These limits were chosen because they represent the bulk of the measured activity for our AGN sample. In this section, we will outline the results of this calculation and explain their significance.\\ 

It is encouraging to see that the top and middle panels, exploring the AGN fraction with stellar mass and redshift respectively, highlight similar trends shown in \cite{Birchall22}. In that work we found little change in AGN fraction with stellar mass, averaging around 1\%. In the top panel of figure \ref{fig:Overall_Frac} there is a similarly flat AGN fraction with stellar mass, averaging about 1\% for the quiescent galaxies, and 2\% for star-forming galaxies. 
\cite{Birchall22} also showed that AGN fraction with redshift increased from around 1\% to 10\%. In the middle panel of figure \ref{fig:Overall_Frac} there is also a clear increase in AGN fraction with redshift. Between $z = 0$ and $0.35$, AGN fraction rises from 0.5\% to 4.5\% for quiescent galaxies, and from 1.5\% to 7\% for star-forming objects.\\ 

Splitting the AGN sample by star-forming classification shows that star-forming galaxies have slightly enhanced AGN fractions. However, this enhancement does not appear statistically significant. To check its significance, we calculated the overall fraction of AGN found in star-forming galaxies and compared it to the fraction in quiescent galaxies. The star-forming AGN fraction was found to be enhanced by a factor of 2 at a $ > 3.5\sigma$ significance. Thus there does appear to be a real increase in the incidence of AGN in star-forming galaxies. \cite{Azadi15} observed a similarly sized star-formation-driven enhancement in a different X-ray selected AGN sample out to $z \approx 1.2$. \\

This enhancement of AGN fraction in star-forming galaxies is also reflected in the bottom panel of figure \ref{fig:Overall_Frac}. The AGN fraction rises from 0.7\% to 3.8\% with increasing $\log_{10} (\mathrm{SFR/SFR_{MS}})$. A similar positive correlation has been observed between average black hole accretion rate and SFR in samples at higher redshifts \citep[e.g.][]{Chen13, Yang17, Pouliasis2022}.  \\ 

\begin{figure}
	\centering
     	\includegraphics[width=\columnwidth]{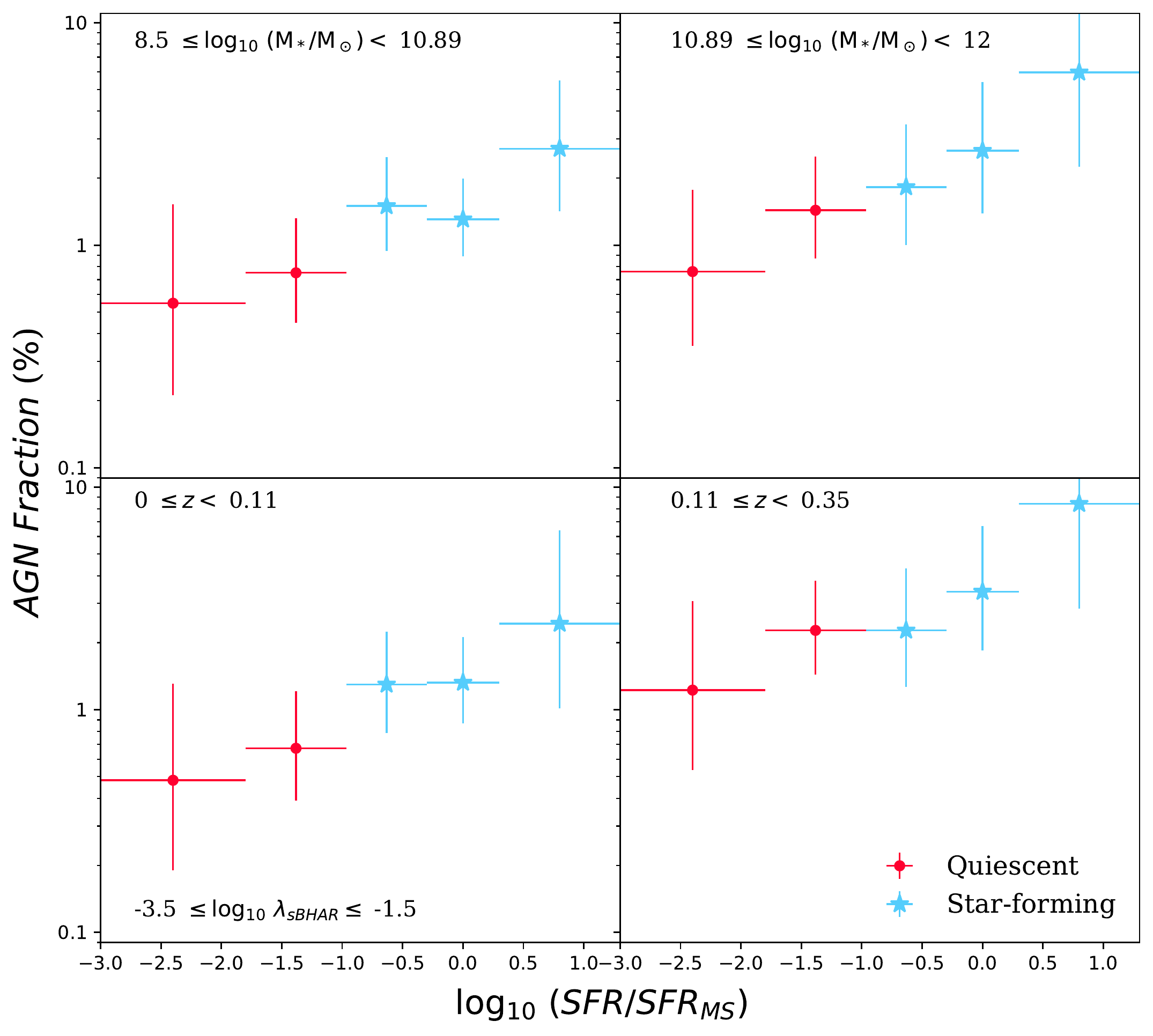}
     	
     	\caption[MPA-JHU AGN fraction as a function of $\log_{10} (\mathrm{SFR/SFR_{MS}})$, split at median stellar mass/redshift]{AGN fraction as a function of $\log_{10} (\mathrm{SFR/SFR_{MS}})$ split at the median stellar mass (top row) and redshift (bottom row) of the sample. In each panel star-forming AGN fractions are shown as blue stars, and quiescent AGN fractions are red circles. The sBHAR ($\lambda_\mathrm{sBHAR}$) integration limits are also displayed in the bottom-left corner.}
	\label{fig:Frac_Rel_SFR_Split}
\end{figure}

Both redshift and $\log_{10} (\mathrm{SFR/SFR_{MS}})$ appear to drive strong increases in AGN fraction. To disentangle these effects, we divided the galaxy and AGN populations and re-plotted the AGN fractions as a function of $\log_{10} (\mathrm{SFR/SFR_{MS}})$ for these new samples. It is clear from figure \ref{fig:Frac_Rel_SFR_Split}, that AGN fraction increases with SFR throughout the sample. Whether split at the median mass ($\log_{10} (M_*/M_\odot) = 10.89$) in the top row or redshift ($z = 0.11$) in the bottom row, both trends show a systematic increase in fraction and a steepening gradient between the low and high value bins.\\

The systematic increase between the low and high redshift bins is expected given the observed trend in figure \ref{fig:Overall_Frac}. Similarly, splitting the sample by mass will shift the average redshift of each bin. So, we would expect an increase in the average fraction in the higher mass driven by the increased redshift. To confirm this, we consolidated each set of fractions from the low and high mass bins into quiescent and star-forming classifications at each panel's median redshift (0.08 for lower mass galaxies, 0.15 for higher mass galaxies). We found that the increase between these consolidated fractions was consistent with the overall AGN fraction increase with redshift.\\

We also see a steepening gradient between low and high value bins which is likely due to changing combinations of star-forming and quiescent hosts. As outlined previously, star formation is known to change with both stellar mass and redshift. So, increasing the average redshift of the bin will change the sample's star-forming properties: the proportion of AGN in star-forming galaxies will increase, those in quiescent galaxies will decrease and produce the observed steeper increase. \\

Whilst this has not been able to precisely disentangle the effects of redshift and $\log_{10} (\mathrm{SFR/SFR_{MS}})$, we have confirmed the effects seen previously. Stellar mass has little effect on the AGN fraction in the nearby Universe, and anything we might see is consistent with a redshift-driven increase. In addition, we have shown that the star-formation-driven enhancement is present at fixed stellar masses and redshifts throughout the nearby Universe. \\

With this analysis, we have shown that the incidence of AGN activity within given accretion limits in the nearby Universe is driven both by the host galaxy's redshift and $\log_{10} (\mathrm{SFR/SFR_{MS}})$. What connects these quantities is the availability of cold gas: the level of star formation is strongly correlated to its availability in a given galaxy, and its abundance is thought to increase at earlier cosmic times \citep[][]{Mullaney12b, Popping12, Vito14}. Thus, our results highlight a strong connection between the availability of cold gas and the level of AGN activity. However, there is still a significant population of AGN within quiescent galaxies. Whilst diminished relative to star-forming galaxies, this population of quiescent AGN implies that another fuelling mechanism is possible, perhaps fuelled by stellar mass loss, as theorised in \cite{KH09}.

\section{Summary \& Conclusions}
\label{sec:Conclusions}
We have investigated how star formation influences the distribution of AGN in the nearby Universe. For this analysis, we created two sets of definitions designed to track changing rates of star formation whilst accounting for redshift and mass driven enhancements. The first definition split the sample into star-forming and quiescent populations; the second, split the sample into even more refined bins of star formation relative to the galactic main sequence. \\

We then applied these star-forming classifications to the BPT-selected AGN sample. Only 72\% of the X-ray selected sample were identified as AGN using the BPT diagnostic. In particular, it is much less effective in quiescent hosts: identifying near all star-forming AGN compared to only 50\% of the X-ray AGN. The other half of the quiescent sample was made up of higher mass galaxies which did not have sufficiently significant emission line activity to make it onto the BPT diagnostic. We believe this is may be due to stellar absorption lines hiding the weaker optical emission lines from low levels of AGN activity. This is further evidence to suggest that the BPT diagram is not effective at identifying weakly accreting AGN, especially in quiescent host galaxies. \\

Next, we investigated how star-forming classification affects the rate at which black holes accrete material. We found that across stellar mass, AGN in star-forming galaxies generally have much higher accretion rates than their quiescent counterparts with a mean difference between star-forming classifications of $\log_{10} \lambda_\mathrm{sBHAR} \sim 0.5$. This effect was most pronounced at the very highest stellar masses ($\log_{10} \mathrm{M_*/M_\odot} \gtrsim 11$).\\  

We built on the probability distribution analysis from \cite{Birchall20, Birchall22} by applying the star-forming classifications to highlight how they affect the AGN population in the nearby Universe. The strongest star-formation-driven changes were seen in the redshift-binned distributions. Quiescent AGN showed a significant steepening with redshift, appearing to favour much lower accretion rate activity. Star-forming galaxies appeared to show the opposite trend with flatter distributions at higher redshifts. This implies star formation facilitates more active accretion at higher redshifts.\\

Finally, by integrating under these distributions we could calculate the fraction of galaxies in the nearby Universe hosting AGN and determine the effect of star formation. Reassuringly, both AGN fraction trends with stellar mass and redshift seen in \cite{Birchall22} are recreated when split by star-forming classification. There is little change in AGN fraction with stellar mass, and a noticeable increase with redshift. We found that star formation increases the incidence of AGN by a factor of 2 at a $> 3.5 \sigma$ significance. This enhancement is also seen when binning the AGN fraction as a function of $\log_{10} (\mathrm{SFR/SFR_{MS}})$. However there is still a significant population of AGN in quiescent galaxies in the nearby Universe. \\ 

In conclusion, we have shown that star formation has an effect on the AGN distribution in the nearby Universe by facilitating higher black hole accretion rates and increasing the probability of finding AGN activity in a given galaxy. However, AGN are still prevalent in quiescent galaxies suggesting additional fuelling mechanisms, such as from stellar mass loss, can also facilitate AGN activity. 

\section*{Acknowledgements}
The authors would like to the reviewer for their helpful comments.
KB acknowledges funding from a Horizon 2020 grant (XMM2Athena). 
JA acknowledges support from an STFC Ernest Rutherford Fellowship (grant code: ST/P004172/1) and a UKRI Future Leaders Fellowship (grant code: MR/T020989/1).
\\

This research has made use of data obtained from the 3XMM XMM-Newton serendipitous source catalogue compiled by the 10 institutes of the XMM-Newton Survey Science Centre selected by ESA.\\

In addition, this research made use of Astropy,\footnote{http://www.astropy.org} a community-developed core Python package for Astronomy \citep{astropy:2013, astropy:2018}.\\

Funding for SDSS-III has been provided by the Alfred P. Sloan Foundation, the Participating Institutions, the National Science Foundation, and the U.S. Department of Energy Office of Science. The SDSS-III web site is \href{http://www.sdss3.org/}{http://www.sdss3.org/}.\\

KB would also like to thank F-X Pineau for his help with constructing the \texttt{xmatch} scripts, and Duncan Law-Green for the help accessing the Flix archive. 


\section*{Data Availability Statement}
The data underlying this article has been derived from publicly available datasets: the MPA-JHU Catalogue (based on SDSS DR8) \& 3XMM DR7. Section \ref{sec:data} outlines where these catalogues are available and how the final sample was derived. The underlying data will also be shared on request to the corresponding author.



\bibliographystyle{mnras}
\bibliography{Bibliography}



\appendix

\onecolumn
\section{Probability Distribution Fit Coefficients}
Table A1 outlines the best-fit coefficients, and associated errors, for equation \eqref{eq:powerlaw} to create every probability distribution shown in figure \ref{fig:Prob_Dist_Comparison}.  

\label{tab:Prob_Dist_Coeffs}

\begin{longtable}{|l|cc|cc|}

\hline

\multicolumn{1}{|c|}{\textit{$\log_{10} x' = -2.55$}} & \multicolumn{2}{c|}{Quiescent AGN}
& \multicolumn{2}{c|}{Star-forming AGN} \\

\hline

$\log_{10}$ Stellar Mass ($\mathrm{M_\odot}$) & $\log_{10}\ A$ & $k$ & $\log_{10}\ A$ & $k$ \\

\hline

8.00 - 9.00 & $-2.44 \pm 1.27$ & $-0.52 \pm 0.04$ & $-2.36 \pm 0.33$ & $-0.98 \pm 0.58$  \\

\rowcolor{Gray}
9.50 - 10.00 & $-2.77 \pm 0.80$ & $-0.91 \pm 0.25$ & $-2.33 \pm 0.31$ & $-0.93 \pm 0.37$   \\

10.00 - 10.50 & $-2.60 \pm 0.23$ & $-0.55 \pm 0.36$ & $-2.36 \pm 0.14$ & $-0.57 \pm 0.19$  \\

\rowcolor{Gray}
10.50 - 11.00 & $-2.50 \pm 0.11$ & $-0.68 \pm 0.15$ & $-2.03 \pm 0.07$ & $-0.48 \pm 0.11$ \\
    
11.00 - 11.50 & $-2.62 \pm 0.14$ & $-1.04 \pm 0.15$ & $-2.14 \pm 0.15$ & $-0.90 \pm 0.21$  \\
\rowcolor{Gray}
11.50 - 12.00 & $-2.53 \pm 0.00$ & $-1.28 \pm 0.02$ & $-1.82 \pm 0.14$ & \ $0.00 \pm 1.36$  \\

\hline

z & $\log_{10}\ A$ & $k$ & $\log_{10}\ A$ & $k$ \\

\hline
\rowcolor{Gray}
0 - 0.05 & $-2.64 \pm 0.27$ & $-0.59 \pm 0.27$ & $-2.29 \pm 0.13$ & $-0.66 \pm 0.17$  \\

0.05 - 0.10 & $-2.63 \pm 0.15$ & $-0.79 \pm 0.17$ & $-2.37 \pm 0.11$ & $-0.85 \pm 0.15$  \\
\rowcolor{Gray}
0.10 - 0.15 & $-2.51 \pm 0.15$ & $-1.16 \pm 0.16$ & $-2.18 \pm 0.15$ & $-1.02 \pm 0.26$  \\

0.15 - 0.20 & $-2.51 \pm 0.20$ & $-1.31 \pm 0.25$ & $-2.02 \pm 0.15$ & $-0.96 \pm 0.30$  \\
\rowcolor{Gray}
0.20 - 0.25 & $-2.36 \pm 0.30$ & $-1.32 \pm 0.36$ & $-1.71 \pm 0.23$ & $-0.66 \pm 0.55$  \\

0.25 - 0.30 & $-2.57 \pm 0.00$ & $-1.99 \pm 0.01$ & $-1.50 \pm 0.27$ & $-0.51 \pm 0.60$  \\

\hline

$\log_{10} \mathrm{SFR/SFR_{MS}}$ & $\log_{10}\ A$ & $k$ & $\log_{10}\ A$ & $k$ \\

\hline
\rowcolor{Gray}
-3.00 - (-)1.80 & $-2.79 \pm 0.15$ & $-1.10 \pm 0.15$ &  &   \\

-1.80 - (-)0.965 & $-2.42 \pm 0.09$ & $-0.73 \pm 0.11$ &  &   \\
\rowcolor{Gray}
-0.965 - (-)0.3 & & & $-2.22 \pm 0.09$ & $-0.64 \pm 0.12$ \\

-0.30 - 0.30 & & & $-2.24 \pm 0.08$ & $-0.86 \pm 0.13$ \\
\rowcolor{Gray}
\ 0.30 - 1.30 & & & $-1.95 \pm 0.14$ & $-0.90 \pm 0.27$ \\

\hline

\multicolumn{5}{|l|}{\textbf{Table A1:} Best-fit coefficients used to fit equation \eqref{eq:powerlaw} to all probability distribution configurations.}\\

\hline

\end{longtable}

\section{Other Probability Distributions}

\label{app:Prob_Dists}
\begin{figure*}
    \centering
    \includegraphics[width=\columnwidth]{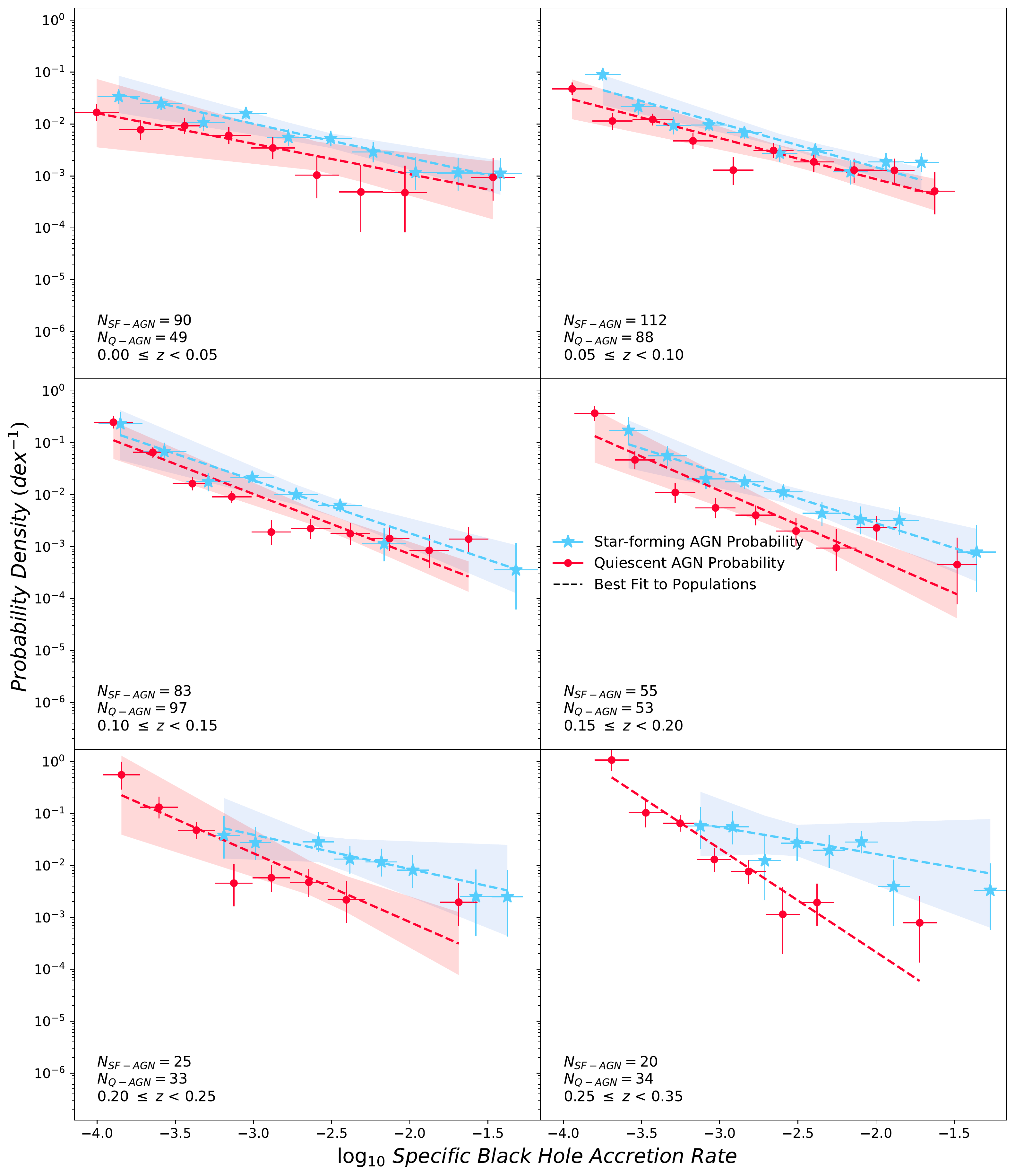}

    \caption[Additional MPA-JHU AGN probability distributions: $\mathrm{\lambda_{sBHAR}}$ \& redshift]{Additional probability distribution used to calculate fits and associated error regions for the other host galaxy properties. This figure shows the probability of a galaxy hosting an AGN as a function of $\mathrm{\lambda_{sBHAR}}$ and split into bins of redshift. The star-forming AGN population is shown as blue stars, and the quiescent AGN population as red circles. As with figure \ref{fig:Example_Prob_Dist}, power laws (dashed lines) have been fit to the data in each panel and displayed alongside their $1\sigma$ uncertainty (pale region). These plots were constructed using the method outlined in section \ref{sub_sec:prob_dist_create}.}
    \label{fig:Other_Prob_Dists_1}
\end{figure*}

\begin{figure*}
    \centering
    \includegraphics[width=0.9\columnwidth]{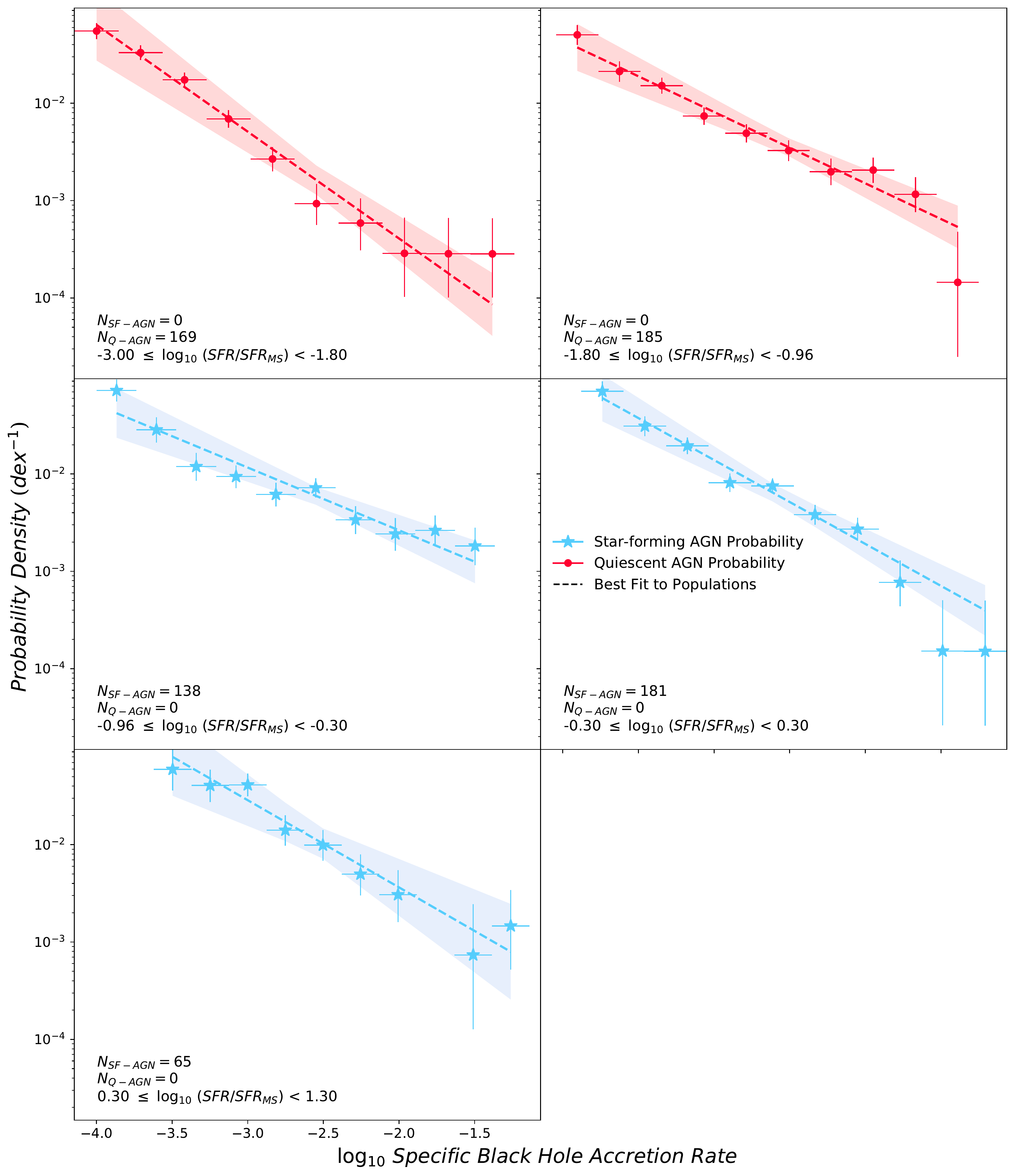}
    
    \caption[Additional MPA-JHU AGN probability distributions: $\mathrm{\lambda_{sBHAR}}$ \& relative SFR]{Additional probability distribution used to calculate fits and associated error regions for the other host galaxy properties. This figure shows the probability of a galaxy hosting an AGN as a function of $\mathrm{\lambda_{sBHAR}}$ and split into bins of $\log_{10} \mathrm{(SFR/SFR_{MS})}$ (see (§\ref{subsec:SFR_rel_MS} for more information on how this quantity was calculated). The star-forming AGN population is shown as blue stars, and the quiescent AGN population as red circles. As with figure \ref{fig:Example_Prob_Dist}, power laws (dashed lines) have been fit to the data in each panel and displayed alongside their $1\sigma$ uncertainty (pale region). These plots were constructed using the method outlined in section \ref{sub_sec:prob_dist_create}.}
    \label{fig:Other_Prob_Dists_2}
\end{figure*}

Figures \ref{fig:Other_Prob_Dists_1} and \ref{fig:Other_Prob_Dists_2} show the probability distributions for star-forming and quiescent AGN, split by redshift, and $\log_{10} \mathrm{(SFR/SFR_{MS})}$. The remaining trends with sBHAR are presented for the sake of transparency, to show the strength of our power law fits to the data. All the plots have the same form as figure \ref{fig:Example_Prob_Dist}.

\bsp	
\label{lastpage}
\end{document}